\documentclass[conference]{IEEEtran}
\IEEEoverridecommandlockouts
\usepackage{cite}
\usepackage{amsmath,amssymb,amsfonts}
\usepackage{algorithmic}
\usepackage{subcaption}
\usepackage{enumitem}
\usepackage{booktabs}
\usepackage{tabularray}
\usepackage{threeparttable}
\usepackage{marvosym}
\usepackage{csquotes}
\usepackage{enumitem}
\usepackage{color}
\usepackage{amsmath}
\allowdisplaybreaks 
\usepackage{makecell}
\usepackage{multirow}
\usepackage{float}
\usepackage{colortbl}
\usepackage{hhline}
\usepackage[most]{tcolorbox}  
\usepackage{balance}
\usepackage[normalem]{ulem}
\usepackage{url}

\usepackage{lipsum}
\usepackage[most]{tcolorbox} %
\usepackage[absolute,overlay]{textpos}

\newcounter{rq}

\newtcolorbox[use counter=rq]{objectivebox}[1][]{
  enhanced,
  breakable,
  colback=blue!3,
  colframe=blue!80!black,
  colbacktitle=blue!80!black,
  coltitle=white,
  fonttitle=\bfseries,
  attach boxed title to top left={xshift=2mm,yshift*=-2mm},
  boxed title style={
    rounded corners,
    arc=2mm,
    boxrule=0pt,
  },
  boxrule=1pt,
  arc=1.5mm,  
  boxsep=2pt, 
  left=4pt,   
  right=4pt,
  top=3pt,
  bottom=3pt,
  before skip=0.7em,
  after skip=0.9em,
  title={Objective},
  #1
}

\newtcolorbox{takeawaybox}[1][]{
  enhanced,
  breakable,
  colback=green!3,
  colframe=green!45!black,
  colbacktitle=green!45!black,
  coltitle=white,
  fonttitle=\bfseries,
  attach boxed title to top left={xshift=2mm,yshift*=-2mm},
  boxed title style={
    rounded corners,
    arc=2mm,
    boxrule=0pt,
  },
  boxrule=1pt,
  arc=1.5mm,
  boxsep=2pt,
  left=4pt,
  right=4pt,
  top=3pt,
  bottom=3pt,
  before skip=0.7em,
  after skip=0.9em,
  title={Finding},
  #1
}

\usepackage{array}

\usepackage{graphicx}
\definecolor{Silver}{rgb}{0.819,0.819,0.819}

\usepackage{siunitx}
\usepackage{graphicx}
\usepackage{textcomp}
\usepackage[dvipsnames]{xcolor}

\def\BibTeX{{\rm B\kern-.05em{\sc i\kern-.025em b}\kern-.1em
T\kern-.1667em\lower.7ex\hbox{E}\kern-.125emX}}

\newcommand{\tight}{\hspace{4pt}} %

\begin{document}
\title{\makebox[\linewidth][c]{Predicting Lakehouse Performance in Clouds:}\\
\makebox[\linewidth][c]{An Empirical Exploration of Query Runtime Variance}
}

\author{\IEEEauthorblockN{James Nurdin}
\IEEEauthorblockA{University of Glasgow \\
Glasgow, United Kingdom \\
j.nurdin.1@research.gla.ac.uk}
\and
\IEEEauthorblockN{Wei Liu}
\IEEEauthorblockA{Barclays PLC \\
Glasgow, United Kingdom \\
wei.liu@barclays.com}
\and 
\IEEEauthorblockN{Richard Mccreadie}
\IEEEauthorblockA{University of Glasgow \\
Glasgow, United Kingdom \\
richard.mccreadie@glasgow.ac.uk}
\and 
\IEEEauthorblockN{Lauritz Thamsen}
\IEEEauthorblockA{University of Glasgow \\
Glasgow, United Kingdom \\
lauritz.thamsen@glasgow.ac.uk}
}

\maketitle

\begin{textblock*}{\textwidth}(17.0mm,260mm)
    \begin{tcolorbox}[width=\textwidth, colback=gray!10, colframe=gray!10, sharp corners, boxrule=0.5pt, boxsep=2pt]
        \centering \footnotesize \textbf{For the purpose of open access, we have applied a Creative Commons Attribution (CC BY) license to this version of our paper.}
    \end{tcolorbox}
\end{textblock*}

\begin{abstract}
Data analytics increasingly runs on distributed lakehouse systems, where platform operators must optimise monetary, resource, and environmental costs. Query Performance Prediction (QPP) helps to balance these costs and supports workload management techniques, such as adaptive resource scaling and low-carbon scheduling. However, runtimes in lakehouses can vary substantially, and the impact of runtime variance on QPP accuracy and workload orchestration has not previously been systematically studied for lakehouse systems.

\looseness=-1 This paper addresses this gap by investigating the runtime variance observed for distributed lakehouse analytical queries and its impact on QPP. First, we quantify the run-to-run variance using Kubernetes deployments across three public clouds and one private cloud, spanning multiple database scales and three analytical benchmarks. Our results demonstrate that repeated executions of the same query can vary in runtime by nearly twofold. Second, we conduct a factor analysis study assessing key sources of this runtime variance such as data locality, co-tenant load, and caching effects. Third, we examine how variance influences state-of-the-art QPP models, revealing that addressing key sources of variance can reduce prediction error up to 80\%. Finally, we demonstrate the downstream implications for low-carbon scheduling as an example of a workload management technique that relies on performance prediction, showing that accounting for runtime variance can lead to a significant reduction in carbon costs.
\end{abstract}

\section{Introduction}
Modern data warehouse platforms such as Google BigQuery, AWS Redshift, and Azure Synapse execute query workloads on cloud infrastructure to support reporting and decision-making. As data volumes and workloads have grown, platform operators have increasingly adopted data lakehouses, which unify the ACID guarantees of data warehouses with the scalability of data lakes to enable SQL analytics over heterogeneous data through shared metadata and table schemas~\cite{9597091,10020719,armbrust2021lakehouse,jain2023analyzing,camacho2024lst,harby2025data, zhang2025precomputation}.

The operation of distributed analytics platforms generates costs, regardless of whether the systems run on private or public clouds. On-premise deployments require investment in infrastructure and come with costs for energy consumption, while public cloud resources incur pay-per-use charges for virtual compute and storage resources. In both cases, there are also environmental costs arising from associated carbon emissions. One approach to actively managing these costs is leveraging query performance prediction (QPP) for optimised workload orchestration. Typically, QPP models are trained on historic executions and used to anticipate runtimes before execution~\cite{4550823, 6228100, Hilprecht2022a,MarcusQPPNet, twitter, microsoft}, which can inform orchestration tasks such as admission control, query optimisation, and scheduling.

\looseness=-1Runtime variance is known to limit the accuracy of performance prediction for distributed data analytics in cloud environments~\cite{10.14778/2733085.2733092,10.14778/1687627.1687707, 10.1145/3185768.3186299, 8241102, 10.1145/3588921}.
However, despite the growing adoption of lakehouse systems, the extent and implications of runtime variance in this context remain insufficiently examined and understood. We argue that runtime variance in distributed lakehouses is too significant to be treated as negligible background noise and that it needs to be accounted for in prediction-driven workload orchestration in both public and private clouds. To support this, we conduct a series of studies that first characterise run-to-run variance by repeatedly executing analytical workloads on Trino-based deployments across three public clouds (GCP, AWS, Azure) and an on-premise, private cloud. We then examine key sources of variance through controlled adjustments to deployment conditions, quantify how variance propagates into QPP error, and illustrate downstream implications using low-carbon query scheduling as a prediction-driven workload optimisation task. Overall, this paper provides the first systematic evaluation of runtime variance in distributed lakehouses, its impact on QPP, and its implications for prediction-driven workload optimisation. Our contributions are:
\begin{enumerate}
\item \textbf{Empirical characterisation:} We characterise observed run-to-run variance in distributed lakehouses across deployment platforms and workload scales (Section~\ref{Study1}).
\item \textbf{Factor analysis:} We isolate the impact of key sources of runtime variance through controlled private-cloud and public-cloud experiments (Section~\ref{Study2}).
\item \textbf{QPP examination:} We evaluate sources of variance by training and testing representative QPP models on historical query executions, demonstrating that reducing runtime variance leads to improved QPP accuracy (Section~\ref{Study3}). 
\item \textbf{Scheduling implications:} Using low-carbon query scheduling, we show that runtime variance affects downstream cost optimisation and that accounting for this uncertainty can reduce observed emissions (Section~\ref{Study4}).
\item \textbf{Artifacts:} We release Helm manifests, workloads, scripts, and analysis documentation to support reproduction\footnote{\url{https://github.com/GlasgowC3lab/lakehouse-variance}}.
\end{enumerate}

\section{Related Work}\label{sec:related_work}   
In this section, we first review the work on runtime variance in distributed analytics. We then discuss runtime QPP for distributed analytics and low-carbon scheduling, highlighting how gaps in the literature motivate our research.
\enlargethispage{1\baselineskip}
\subsection{Runtime Variance in Database and Analytics Systems}
Runtime variance has been extensively studied in distributed database systems, with works recognising variance as a crucial factor for both performance and reliability. Zhu et al.~\cite{10.1145/3588921} investigate the runtimes of historical analytics jobs on Microsoft’s Cosmos platform, attributing variance between repeated jobs to resource allocation, cluster heterogeneity, and scheduling policies. Schad et al.~\cite{10.14778/1920841.1920902} study runtime variance in public cloud infrastructures using a distributed MapReduce workload, showing that performance on Amazon EC2 varies substantially and identifying VM type, AZ, and time-of-day effects as key contributors. Huang et al.~\cite{huang2016identifying} introduce VProfiler, a runtime variance profiler for DBMSs that identifies sources of uncertainty, revealing lock scheduling and logging as key influences for variance in MySQL and PostgreSQL. Aktaş et al.~\cite{aktacs2019straggler} conduct a cost-latency analysis of variance curtailment mechanisms. By using analytical models driven by Google cluster traces, results show that the tail heaviness of service times critically influences trade-offs between reducing variance or overall latency. Ferreira et al.~\cite{10955378} evaluate how reducing variance impacts the performance of distributed NoSQL databases across cloud deployments, showing with YCSB workloads on Cassandra, MongoDB, and Redis that enforcing consistency can degrade performance by up to 95\%.

Alongside studies on variance, several recent works examine the analytical performance of data lakehouse deployments. Jain et al.~\cite{jain2023analyzing} propose the LHBench benchmark to analyse lakehouse table storage designs and compare how performance varies between storage characteristics. Camacho et al.~\cite{camacho2024lst} introduce the LST-Bench benchmark to evaluate the ACID properties of lakehouses by measuring performance under different compression, versioning, and concurrent I/O settings in cloud environments. Harby et al.~\cite{harby2025data} conduct an experimental study comparing a lakehouse deployment with data warehouses and data lakes, measuring ingestion latency and analytical query performance to highlight operational benefits. 

While prior work has extensively explored runtime variance in distributed database systems, these studies focus on traditional DBMSs, MapReduce workloads, and NoSQL systems rather than modern lakehouse architectures. Conversely, recent research provides valuable insight into the performance of lakehouse systems, but has focused primarily on query throughput and latency. To the best of our knowledge, no study investigates sources of runtime variance in lakehouses.
\subsection{Runtime QPP Modelling in Distributed Analytics}

A substantial strand of work investigates QPP in cluster-based and cloud analytics environments. Al-Sayeh et al.\cite{10.1145/3514221.3517892} present Juggler, a learned framework for Spark-based ML workloads that predicts dataset sizes and runtimes to choose cache plans and cluster configurations that fit in memory and reduce both execution time and cost. Scheinert et al.~\cite{10253884} propose Karasu, a collaborative resource-configuration profiler that leverages lightweight performance models shared across users, combining them in a Bayesian ensemble which identifies near-optimal cluster configurations for runtime, cost, and energy. Venkataraman et al.~\cite{194946} introduce Ernest, a black-box performance prediction framework for large-scale analytics that runs jobs on small input samples and uses simple system-level models with optimal experiment design to accurately predict runtimes across data and cluster scales. Song et al.~\cite{song2025linkedinqpp} evaluate several learned QPP models on LinkedIn’s Trino-based workloads, highlighting both the promise and the practical limitations of state-of-the-art models when deployed on real, large-scale production clusters. Complementary to these efforts, Strausz et al.~\cite{Strausz2025CrossEngineOptimizerCDMS} propose a learned cost-model-based cross-engine optimiser deployed in a data lakehouse, using a graph neural network to predict execution time across multiple engines and deployments and to route queries accordingly.

\looseness=-1Across these studies, results show accurate runtime and cost prediction is achievable for distributed analytics systems, however most QPP models target average latency or cost and treat run-to-run variance as noise. Even in recent lakehouse work using learned cost models, the emphasis is on engine selection rather than how variance limits QPP reliability, if it is not actively accounted for. 
\enlargethispage{1\baselineskip}
\vspace{-4pt}
\subsection{Low-Carbon Scheduling}
\vspace{-3pt}
\looseness=-2Finally, recent work has increasingly explored the area of low-carbon scheduling, which investigates how resource management systems can leverage carbon-intensity (CI) signals for optimisation, shifting flexible workloads toward times or regions with better sustainable energy availability rather than optimising solely for throughput. Radovanović et al.~\cite{9770383} introduce Google’s Carbon-Intelligent Compute Management System, which uses day-ahead workload and carbon-intensity forecasts to optimise virtual capacity curves for flexible data centre workloads, reporting a 1--2\% reduction in power consumption during higher CI periods. By analysing regional CI patterns, Wiesner et al.~\cite{Wiesner2021LetsWA} simulate scheduling policies for delay-tolerant workloads.  Their results demonstrate that deferring jobs toward cleaner energy periods can reduce emissions by 5\% across all examined regions. Bahreini et al.~\cite{bahreini2023carbon} propose a low-carbon workload dispatcher across geographically distributed clouds. By leveraging linear-programming, their joint job placement and scheduling algorithm achieves near-optimal schedules with approximation ratios between 1.1 and 1.2. More recently, Rodrigues et al.~\cite{rodrigues2025carbon} propose LinTS, a carbon-aware data scheduler for inter-data centre data transfers. By considering temporal and scaling decisions, they report reductions in carbon emissions by up to 66\% in worst case scenarios and by 15\% compared with previous heuristic schedulers.

Collectively, these studies establish low-carbon query scheduling as a relevant optimisation task across cloud platforms. However, the existing works assume that workload runtime is known or can be predicted reliably, presenting a gap to explore how runtime variance and prediction errors affect the quality of resulting schedules.

\section{Experimental Design}\label{sec:ExperimentalDesign }
In this section, we outline the experimental setup, including the lakehouse architecture, cluster resources, workloads, and key configuration choices used in this paper.

\subsection{Lakehouse Architecture}
\begin{figure}[!h]
\centering
\includegraphics[width=0.94\linewidth]{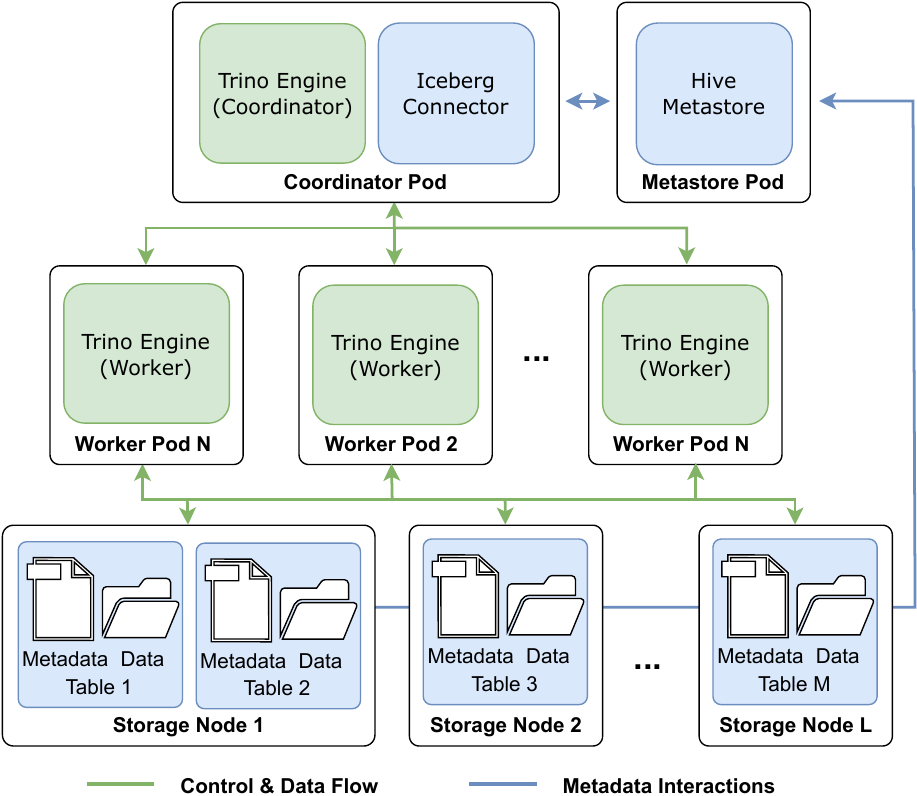}
\caption{Lakehouse architecture overview.}
\label{fig:arch}
\end{figure}

\looseness=-1In a similar way to prior work~\cite{10020719,9597091,armbrust2021lakehouse}, we consider a distributed data lakehouse as a system comprising object storage, a metadata store, an open table format, and a distributed query engine, deployed as services across cluster infrastructure. We implement the lakehouse architecture illustrated in Figure~\ref{fig:arch}, in which Trino serves as the distributed query engine, Apache Iceberg provides the open-table format, and a Hive Metastore facilitates access to distributed object storage. These components are deployed as containerised services orchestrated by Kubernetes.

\looseness=-1To ensure consistent lakehouse settings across deployments, we deploy static container images for required services. Likewise, Trino and Hive Metastore configuration files are kept constant across platforms where possible, with changes limited to authentication settings required for compatibility with the corresponding object storage, following Trino documentation\footnote{\url{https://trino.io/docs/current/connector/iceberg}}. 

\subsection{Cloud Clusters}

We deploy the lakehouse on commodity public-cloud and private-cloud clusters to examine runtime behaviour across different deployment environments. For the public cloud, we consider AWS, Azure, and GCP, while the private-cloud setting provides a contrasting self-managed environment backed by CephFS storage. Instance types used for the public-cloud platforms are summarised in Table~\ref{tab:instances}. For the Private-Cloud cluster, we rely on two CephFS-backed node types (\texttt{LOCAL\_A} and \texttt{LOCAL\_B}), each equipped with an AMD Threadripper Pro 5965WX and a Ryzen 9 7950X CPU, respectively. Across all lakehouse deployments, consistent service-level vCPU and RAM allocations are enforced, with coordinator, worker, and metastore resources provisioned according to Table~\ref{tab:instances}.

\begin{table}[h]
\centering
\caption{Node Provisioning Over Cloud Platforms}\vspace{-0.5em}
\label{tab:instances}
\begingroup
\small
\renewcommand{\arraystretch}{1}
\begin{tabular}{l@{\hspace{3pt}}c@{\hspace{5pt}}c@{\hspace{5pt}}
c@{\hspace{5pt}}c@{\hspace{5pt}}c
}
\toprule
\multirow{2}{*}{\raisebox{-3.4ex}{\makecell[c]{Lakehouse \\Service}}} 
  & \multirow{2}{*}{\raisebox{-3.4ex}{\makecell[c]{vCPUs}}}
  &\multirow{2}{*}{\raisebox{-3.4ex}{\makecell[c]{RAM \\ (GB)}}} 
  &\multicolumn{3}{c}{Cloud Platform (AZ)} \\
\cmidrule(lr){4-6}
& & & \makecell[c]{AWS\\(eu-west-2)} 
& \makecell[c]{Azure\\(uksouth)}
& \makecell[c]{GCP\\(europe-west2)} \\
\midrule
Coordinator
  & 4
  & 16
  & t3a.xlarge 
  & D4as\_v6 
  & e2-standard-4 \\
Worker
  & 4
  & 64
  & r6i.2xlarge 
  & E8s\_v6 
  & e2-highmem-8 \\
Metastore
  & 2
  & 4
  & t3.medium 
  & D4als\_v6 
  & e2-custom \\
\bottomrule
\end{tabular}
\endgroup
\vspace{-1em}
\end{table}

\begin{table*}[!h]
\centering
\footnotesize
\caption{Variance Characterisation Across Lakehouse Deployments}\vspace{-0.5em}
\label{tab:variance-deployments}
\resizebox{0.93\linewidth}{!}{%
\renewcommand{\arraystretch}{0.9}
\begin{tabular}{%
c@{\tight}cc@{\hspace{-1.25em}} %
  S[table-format=3.3]   S[table-format=2.3]
  S[table-format=1.3]
  S[table-format=3.3]
  S[table-format=2.3]
  S[table-format=2.3]
  S[table-format=2.3]
}
\toprule
\multicolumn{2}{c}{\multirow{2}{*}{\raisebox{-0.9ex}{Lakehouse Context}}}& \multicolumn{2}{c}{\multirow{2}{*}{\begin{tabular}[c]{c}\raisebox{-0.5ex}{Mean Query Runtime}\\\raisebox{-0.5ex}{Across Workloads (s)}\end{tabular}}} & \multicolumn{6}{c}{Variance Across 5 Repeated Runs}\\
\cmidrule(lr){5-10} 
\multicolumn{2}{c}{}& \multicolumn{2}{c}{} & \multicolumn{3}{c}{Std. (s)} & \multicolumn{3}{c}{CV (\%)} \\ 
\cmidrule(lr){1-2} \cmidrule(lr){3-4} \cmidrule(lr){5-7} \cmidrule(lr){8-10}
\makecell{Cluster\\Platform} & \makecell{TPC-DS \\ SF} &  \multicolumn{1}{c}{~~~~~Avg.}& \multicolumn{1}{c}{Std.}& \multicolumn{1}{c}{Avg.}& \multicolumn{1}{c}{Med.}& \multicolumn{1}{c}{P99}& \multicolumn{1}{c}{Avg.}& \multicolumn{1}{c}{Med.}& \multicolumn{1}{c}{P99} \\
\midrule
AWS & \multirow{4}{*}{10} &  5.589 & 0.060 & 0.304 & 0.215 & 1.429 & 7.944 & 6.633 & 31.356 \\ %
Azure &&  3.739 & 0.034 & 0.176 & 0.134 & 0.668 & 8.430 & 7.312 & 24.642 \\ %
GCP &&  6.893 & 0.111 & 0.415 & 0.272 & 2.814 & 7.012 & 5.929 & 22.378 \\ %
Private-Cloud &&  3.314 & 0.345 & 0.538 & 0.233 & 4.267 & 20.029 & 17.686 & 76.848 \\ %
\cmidrule(){1-10}
AWS & \multirow{4}{*}{100} &  13.067 & 0.086 & 0.538 & 0.257 & 4.065 & 6.212 & 4.260 & 23.967 \\ %
Azure &&  10.579 & 0.045 & 0.407 & 0.231 & 2.887 & 9.069 & 4.707 & 52.269 \\ %
GCP &&  17.612 & 0.355 & 1.123 & 0.507 & 5.690 & 8.319 & 5.458 & 40.658 \\ %
Private-Cloud &&  17.596 & 4.295 & 5.104 & 1.587 & 72.227 & \textbf{29.560} & \textbf{27.677} & 54.990 \\ %
\cmidrule(){1-10}
AWS & \multirow{4}{*}{1000} &  43.964 & 6.325 & 6.418 & 0.496 & 73.611 & 6.349 & 3.217 & 32.594 \\ %
Azure && 39.755 & 3.466 & 4.860 & 0.855 & 39.002 & 10.520 & 4.819 & 51.134 \\ %
GCP &&  27.999 & 1.696 & 3.087 & 1.103 & 27.958 & 10.835 & 8.545 & 41.058 \\ %
Private-Cloud &&  \textbf{109.432} & \hspace{5pt}\textbf{12.352} & \textbf{19.033} & \textbf{11.995} & \textbf{142.283} & 21.634 & 17.895 & \textbf{98.652} \\ %
\bottomrule
\end{tabular}
}

\end{table*}

\subsection{Workloads}
\looseness=-2We first employ the decision-support benchmark TPC-DS~\cite{tpcds_v4_spec}. Using the accompanying data generator, we evaluate the 100-query workload at three database scale factors (SF~10, SF~100, and SF~1000), corresponding to 10\,GB, 100\,GB, and 1\,TB datasets. For these scales, we deploy 1, 2, and 4 Trino workers, respectively, as the default configurations across all studies, with Study~1 additionally examining the effect of scaling to higher worker counts. These configurations were chosen to support comparability across scales while providing sufficient aggregate memory for workload execution,  with total worker memory increasing up to 256\,GB at the largest scale.
Furthermore, we broaden the evaluation and employ two additional benchmarks, SSB~\cite{o2009star} and JOB~\cite{leis2015good}, to diversify workload characteristics in Study 3. TPC-DS and SSB represent decision-support workloads dominated by scans, aggregations, and filters, whereas JOB is a join-heavy workload.

\section{Study~1: Variance Characterisation}
\label{Study1}
In this first study, we establish a macro view of query runtime variance in distributed lakehouses by examining how variance propagates across cloud platforms and scales.
\begin{objectivebox}
Observe the extent of runtime variance across lakehouse deployments, and quantify how variance manifests across platforms and scales.
\end{objectivebox}

\subsection{Methodology}
\looseness=-1To quantify variance in lakehouse deployments, we consider a set of fixed experimental settings $s \in S$ spanning the four cluster platforms (AWS, Azure, GCP, and Private-Cloud), along with the three TPC-DS dataset sizes. Cloud deployments run on dedicated clusters of virtual machines, whereas the Private-Cloud deployment executes on a shared co-tenant cluster. For each setting, we execute the SQL workload \(W\) a total of \( R\!=\!5 \) times. Each repetition corresponds to executing the full workload under identical deployment conditions, allowing runtime variance to be observed at the query level and aggregated across the workload. On each pass \(r \in \{1,\dots,R\}\), a new lakehouse instance is deployed with each query \(q \in W\) being executed once, yielding a runtime of \(t_{qrs}\). We calculate the following: 
\vspace{0.35\baselineskip}
\newline
\noindent\textbf{Across-workload metrics:} For a given setting $s$, at each repetition $r$ of the workload $W$, performance of the current run is calculated via the mean runtime $\bar{T}_{s,r}$ across queries:
\begin{equation}
\hspace{-2.7em}\bar{T}_{s,r}\; = \frac{1}{|W|}\sum_{q\in W} t_{qrs}
\end{equation}
We report the average (Avg.) and Bessel-corrected standard deviation (Std.) of mean runtimes across repeated runs.
\vspace{0.25\baselineskip}
\newline
\noindent\textbf{Across-query metrics:}
For each $(q,s)$ pair, we obtain a sample distribution consisting of $R$ runtimes $\{t_{q1s},\ldots,t_{qRs}\}$. Using this, we compute the mean $\mu_{qs}$, the Bessel-corrected standard deviation $\sigma_{qs}$ (reported as Std.), and the coefficient of variation $CV_{qs}$ for each sample:

\vspace{-0.65em}
\begin{subequations}
\begin{align}
\mu_{qs}\;&\;=\; \frac{1}{R}\sum_{r=1}^{R} t_{qrs} \label{eq:mu} \\[0.25em]
\sigma_{qs}\; &\;=\; \sqrt{\frac{1}{R-1}\sum_{r=1}^{R}(t_{qrs}-\mu_{qs})^2} \label{eq:sigma} \\[0.25em]
CV_{qs}\; (\text{\%}) &\;=\; \frac{\sigma_{qs}}{\mu_{qs}} \times 100 \label{eq:cv}
\end{align}
\end{subequations}

To quantify the variance across repeated query runtimes, we aggregate across $q \in W$ and report the average (Avg.), median (Med.), and 99th percentile (P99) of both Std. and CV. By reporting distributional summaries across all queries and repeated runs, we summarise both typical runtime variance and the most variable instances observed within each workload.

\subsection{Results}\label{sec:resultsA}
\looseness=-1Table \ref{tab:variance-deployments} presents the runtime performance across the 12 lakehouse contexts. As expected, the average mean query runtime $\bar{T}_{s,r}$ increases with dataset scale. At SF~10, we observe average runtime latencies consistently range from 3--7s across platforms, when scaled to SF~1000, we see latencies on average increase to 28--44s on the public cloud instances and to 109s on the private-cloud. Within each scale factor, differences between cloud deployments are modest, with average mean runtimes showing slight variation across repeated runs. However, on the private-cloud, we see a clear slowdown as the dataset grows in scale. When examining query latencies across repeated workload passes, we find that public cloud lakehouses exhibit moderate variance, with median CVs between 4--9\% and tail-end CVs (P99) between 20--50\%, indicating a clear presence of variance across repeated query executions. In the private-cloud settings, variance is consistently higher, with average CVs exceeding 20\% in most configurations and P99 CV approaching 100\% at SF~1000, revealing substantial runtime variability. The corresponding standard deviations follow a similar pattern, with private-cloud’s P99 standard deviation exceeding 140s at SF~1000.

\begin{takeawaybox}
Across platforms and scales, runtime variance is a common and persistent property of query execution in distributed lakehouses, with CV values reaching up to 98\%.
\end{takeawaybox}
Results from the study demonstrate that lakehouses across all platforms exhibit run-to-run variance, with private-cloud deployments showing the greatest variability and tail-end variance compared to public cloud platforms. One plausible explanation is that the public-cloud deployments run lakehouse workers in virtual machines with stronger resource isolation, whereas the private-cloud deployment runs workers as containers on shared hosts. As a result, repeated executions in the private-cloud may be more exposed to interference from co-tenant workloads competing for CPU, network, and storage resources, which contribute to the greater tail-end variance. Moreover, while mean runtimes across cloud deployments differ only modestly, the observed tail variance indicates that similar average performance can conceal differences in reliability across repeated executions.

\subsection{Worker Scaling Analysis}

\begin{figure}[!h]
\centering
\includegraphics[width=\linewidth]{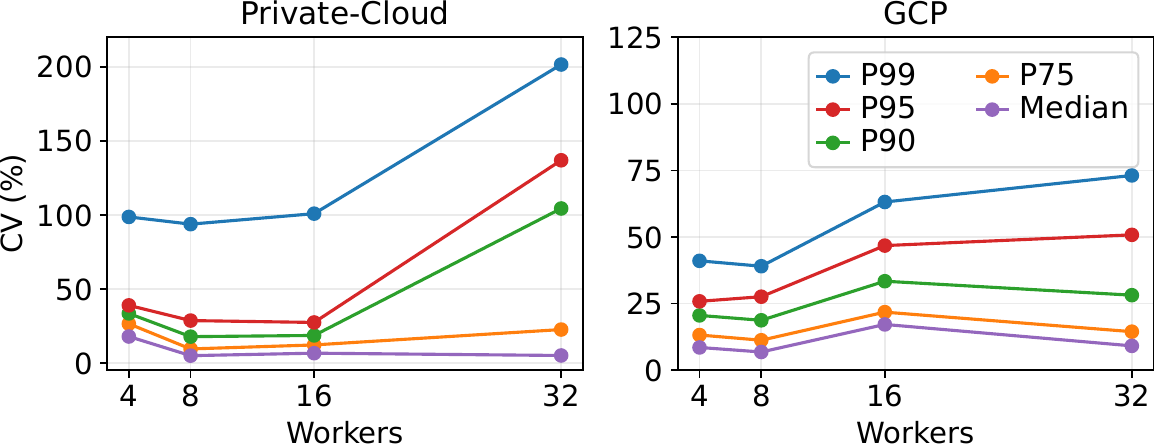}
\caption{Worker-scaling analysis of runtime variance.}
\label{fig:worker-scaling-cv}
\end{figure}

To examine the sensitivity of runtime variance against the total number of Trino workers, we conduct a worker-scaling experiment on both the Private-Cloud and GCP clusters. Executing the TPC-DS workload over the SF~1000 dataset, we begin with the 4-worker deployment from the previous experiment and increase the number of workers while maintaining a fixed memory budget of 256\,GB. Table~\ref{tab:worker-scaling-config} summarises the resource allocation across lakehouse configurations. As the 4-worker deployment already provisions sufficient memory to execute the workload, this analysis focuses on scaling parallel compute through worker counts of 4, 8, 16, and 32. Following the same methodology as before, we compute the CV of query runtimes across 5 repeated runs. Figure~\ref{fig:worker-scaling-cv} presents the observed CV for the different worker configurations.

\begin{table}[H]
\centering
\caption{Resource Allocation Across Worker Scales}\vspace{-0.5em}
\small
\renewcommand{\arraystretch}{0.7}
\begin{tabular}{%
ccccc} 
\toprule
\multirow{2}{*}{\makecell{Worker\\Count}} & \multicolumn{2}{c}{vCPUs} & \multicolumn{2}{c}{RAM\,(GB)}  \\ 
\cmidrule(lr){2-3}\cmidrule(lr){4-5}
 & Per Worker& Total & Per Worker & Total \\ 
\cmidrule(lr){1-5}
4 & 4  & 16   & 64&  256\\
8 & 4  & 32   & 32&  256\\
16 & 4 & 64   & 16&  256\\
32 & 4 & 128 & 8&  256 \\
\bottomrule
\end{tabular}
\label{tab:worker-scaling-config}
\end{table}

On the Private-Cloud, runtime variance remains relatively stable from 4 to 16 workers, but rises sharply at 32 workers, particularly in the tail, suggesting a non-linear relationship in which variance remains consistent up to a threshold before increasing disproportionately at larger worker counts. A plausible explanation for this behaviour is that, as data and intermediate results are exchanged continuously throughout query execution, adding more workers increases the communication among workers, which may lead to network congestion and queueing delays that amplify runtime uncertainty. In contrast, the GCP deployment exhibits only moderate increases in variance, suggesting that the underlying infrastructure sustains more stable communication under higher degrees of parallelism and interactions between deployed workers.

\begin{table*}[tpb]
\footnotesize
\caption{Factor Analysis Results}
\vspace{-0.5em}
\label{tab:factor-runtime}
\resizebox{\linewidth}{!}{%
\begin{threeparttable}
\centering
\begin{tabular}{
@{\hspace{15pt}}@{}cl
S[table-format=3.3]l
S[table-format=2.3]S[table-format=2.3]S[table-format=2.3]
lll
}
\toprule
\multirow{3}{*}{\raisebox{-3.3ex}{\makecell{Cluster \\ Platform}}} 
& \multicolumn{1}{c}{\multirow{3}{*}{\raisebox{-3.3ex}{\makecell{Factor}}}}
&\multicolumn{2}{c}{\multirow{2}{*}{\raisebox{-1.4ex}{\makecell{Mean Query Runtime\\Across Workloads}}}} & \multicolumn{6}{c}{Variance Across 5 Repeated Runs}\\
\cmidrule(lr){5-10}
&& \multicolumn{2}{c}{}
& \multicolumn{3}{c}{CV (\%)}
& \multicolumn{3}{c}{ $\Delta$CV (\%)} \\ 
\cmidrule(lr){3-4} \cmidrule(lr){5-7} \cmidrule(lr){8-10} 
& & \multicolumn{1}{c}{\hspace{8pt}Avg. (s)} & \multicolumn{1}{c}{$\Delta$Avg. (\%)}& \multicolumn{1}{c}{Avg.} & \multicolumn{1}{c}{Med.} & \multicolumn{1}{c}{P99} & \multicolumn{1}{c}{Avg.}& \multicolumn{1}{c}{Med.} & \multicolumn{1}{c}{P99}\\ 
\midrule
\rowcolor{gray!25}
\multirow{5}{*}{Private-Cloud} & Baseline & 109.432 & \multicolumn{1}{c}{\textbf{—}} & 21.634 & 17.895 & 98.652 & \multicolumn{1}{c}{\textbf{—}} & \multicolumn{1}{c}{\textbf{—}} & \multicolumn{1}{c}{\textbf{—}} \\ %
& \textbullet\ Warm cache & 102.856 & $-~6.009$ & 24.115 & 20.796 & 64.355 & $+~11.466$ & $+~16.209$ & $-~34.765$ \\ %
& \textbullet\ Local disk storage & \textbf{65.553} & \boldmath$-~40.097$\unboldmath & 18.182 & 14.022 & \textbf{56.496} & $-~15.957$ & $-~21.643$ & \boldmath$-~42.732$\unboldmath \\ %
& \textbullet\ Reduced Co-tenant load & 133.316 & $+~21.826$ & \textbf{15.790} & 11.752 & 57.528 & \boldmath$-~27.014$\unboldmath & $-~34.328$ & $-~41.686$ \\ %
& \textbullet\ Pinned nodes & 90.392 & $-~17.399$ & 16.338 & \textbf{10.685} & 73.910 & $-~24.483$ & \boldmath$-~40.290$\unboldmath & $-~25.080$ \\ %
\cmidrule(){1-10}
\rowcolor{gray!20}
\multirow{5}{*}{GCP} & Baseline & 27.999 & \multicolumn{1}{c}{\textbf{—}} & 10.835 & 8.545 & 41.058 & \multicolumn{1}{c}{\textbf{—}} & \multicolumn{1}{c}{\textbf{—}} & \multicolumn{1}{c}{\textbf{—}} \\ %
& \textbullet\ Warm cache & 34.955 & $+~24.844$ & 17.797 & 15.057 & 39.370 & $+~64.253$ & $+~76.219$ & $-~4.112$ \\ %
& \textbullet\ Virtual disk storage & 56.927 & $+~103.319$ & 13.664 & 11.816 & 40.793 & $+~26.102$ & $+~38.291$ & $-~0.645$ \\ %
& \textbullet\ Increased Co-tenant load & 43.961 & $+~57.013$ & 16.254 & 13.215 & 47.812 & $+~50.004$ & $+~54.657$ & $+~16.450$ \\ %
& \textbullet\ External metastore & 43.744 & $+~56.238$ & 21.314 & 19.069 & 61.279 & $+~96.712$ & $+~123.168$ & $+~49.250$ \\ %
\bottomrule
\end{tabular}%
\end{threeparttable}
} 
\end{table*}

\section{Study~2: Factor Analysis}
\label{Study2}
To isolate the key contributors to runtime uncertainty, this section investigates major sources of variance in lakehouses across deployment and infrastructure conditions.
\begin{objectivebox}
Identify deployment factors associated with runtime variance and evaluate how controlling them impacts runtime variance across repeated runs.
\end{objectivebox}

\subsection{Methodology}
Considering the TPC-DS workload at SF~1000, we deploy lakehouses on our Private-Cloud cluster and on GCP. Using the lakehouse contexts from Study~1 as baselines, we deploy factor variants by controlling each factor $f$ independently. On Private-Cloud, we have full access to the underlying hardware, enabling us to study factors that require low-level monitoring and control. In contrast, GCP restricts deployment configurations to those permitted by the cloud provider, limiting the factors we can study to platform-supported options. We consider the following factors:
\newline

\noindent\textit{Private Cloud factors:}
\begin{itemize}
  \item \textbf{Warm cache}: To understand how cold-start effects influence variance, at each pass $r$ we sequentially execute each query twice and consider runtimes from the second execution, where table data is already cached on workers.
  \item \textbf{Local disk storage}: To attribute variance associated with input data locality in distributed storage, we materialise an additional copy of each table on the ephemeral storage of every Private-Cloud node and update lakehouse table locations to scan local replicas instead of the CephFS.
  \item \textbf{Reduced co-tenant load}: To assess how variance is induced by co-tenant activity, we obtain runtime labels from an evaluation window in which the average CPU utilisation across Private-Cloud nodes is approximately $30\%$ lower than during the baseline period.
  \item \textbf{Pinned nodes}: To study how lakehouse placement affects variance, we pin the coordinator and workers to a fixed subset of \texttt{LOCAL\_A} nodes using node-affinity rules, preventing random node scheduling across passes $r$.
\end{itemize}
\noindent\textit{Cloud platform factors (GCP):}
\begin{itemize}
  \item \textbf{Warm cache}: To assess the role of caching under managed cloud deployments, we apply the same approach as on Private-Cloud, executing each query twice per pass $r$ and using runtimes from the warm executions.
  \item \textbf{Virtual disk storage}: To study how cloud storage influences variance, we replicate table data onto attached virtual disks on each GCP node and redirect table scans to this mounted storage rather than the distributed store.
  \item \textbf{Increased co-tenant load}: To examine sensitivity to co-tenant interference, we deploy the lakehouse alongside the CloudSuite In-Memory Analytics benchmark~\cite{10.1145/2248487.2150982}, configured to consume approximately $30\%$ of each node's resources to mimic a co-tenant load.
  \item \textbf{External metastore}: To consider vendor-offered metadata services, we replace the locally deployed Hive Metastore with a Dataproc Metastore instance in the same region and register all tables with this external service.
\end{itemize}

As in the previous study, for each applied factor $f$, we execute the workload \( R\!=\!5 \) times and obtain the per-query runtime samples $\{t_{q1s},\ldots,t_{qRs}\}$. We report metrics relative to baseline deployments, including the mean query runtime $\bar{T}_{s,r}$ and the Avg., Med., and P99 CV of repeated query runtimes.

\subsection{Results}\label{sec:resultsB}

Table~\ref{tab:factor-runtime} reports, for each cluster platform and lakehouse factor, how mean runtime and run-to-run variance (average, median, and P99 CV) change across repeated executions relative to the baseline configuration. In the Private-Cloud cluster, the baseline configuration provides an average runtime of 109s, with an average CV of 22\%, a median CV of 18\%, and a P99 CV approaching 100\%, exhibiting substantial tail variance. With warm cache enabled, mean runtime and P99 CV decrease slightly, while average and median CV increase, indicating a trade-off in which heavy-tailed variance is reduced at the cost of run-to-run consistency. As expected, staging table data on local disks reduces mean runtime by 43\% and decreases average, median, and P99 CV by between 15\% and 40\%. Under reduced co-tenant load, mean runtime increases, contrary to expectations of faster throughput, while average, median, and P99 CV decrease by up to 41\%. As anticipated, pinning lakehouses to a fixed set of nodes improves both runtime and variance metrics, with median CV decreasing by 40\%.\newline

\begin{table*}[h]
\centering
\caption{Effect of Variance-Skewed Training on QPP Accuracy}\vspace{-0.5em}
\label{tab:qpp-variance}
\renewcommand{\arraystretch}{0.94}

\resizebox{\linewidth}{!}{%
\begin{tabular}{
cl
ccc
ccc
ccc
} 
\toprule
\multicolumn{1}{c}{\multirow{2}{*}{\raisebox{-1.3ex}{\makecell{~Model}}}} & \multirow{2}{*}{\raisebox{-1.5ex}{\makecell{Lakehouse Regime}}}
& \multicolumn{3}{c}{TPC-DS}& 
\multicolumn{3}{c}{SSB} & 
\multicolumn{3}{c}{JOB} \\
\cmidrule(lr){3-5} \cmidrule(lr){6-8} \cmidrule(lr){9-11}
&& MAE (s) & Med QErr & P99 QErr & MAE (s) & Med QErr & P99 QErr & MAE (s) & Med QErr & P99 QErr\\ 
\midrule
\multirow{2}{*}{GNN}
& Higher-Variance & 26.849 & 4.324 & 239.796 & 37.839 & 3.785 & 324.906 & 4.241 & 8.547 & 33.741\\
& Lower-Variance & \textbf{13.571} & \textbf{2.902} & \textbf{102.850} & \textbf{7.752} & \textbf{1.982} & \textbf{38.216} & \textbf{3.981} & 8.213 & \textbf{28.503} \\
\cmidrule(lr){1-11}
\multirow{2}{*}{\makecell{RF}}
& Higher-Variance & 27.743 & 3.727 & 283.716 & 43.661 & 2.870 & 429.989 & 4.521 & 6.372 & 63.698\\

& Lower-Variance & 17.403 & 5.643 & 223.606 & 11.082 & 2.424 & 115.591 & 4.395 & \textbf{6.529} & 64.048\\
\bottomrule
\end{tabular}
} \vspace{-1em}
\end{table*}

\looseness=-1In the GCP cluster, the baseline configuration results in an average runtime of 28s, with average CV of 11\%, median CV of 9\%, and P99 CV of 41\%, suggesting a comparatively stable deployment. Enabling warm cache unexpectedly increases mean runtime to 35s and raises average and median CV by 64\% and 76\%, respectively, in contrast to the Private-Cloud cluster, while P99 CV remains largely unchanged. Employing virtual disk storage more than doubles mean runtime relative to baseline and increases average and median CV, with minimal change in P99 CV. Unsurprisingly, increased co-tenant load raises mean runtime and all three variance metrics by 16--57\%. Utilising an external metastore produces the highest variance among GCP configurations, with average CV at 97\%, median CV at 123\%, and P99 CV at 49\%.

\begin{takeawaybox}[before skip=1em, after skip=1em]
Deployment configurations significantly influence runtime variance: in our Private-Cloud setup, local disk storage reduces variance by up to 43\%, whereas in GCP, external metadata increases variance by up to 123\%.
\end{takeawaybox}
\vspace{5pt}
These results suggest that runtime variance in lakehouses is highly sensitive to deployment decisions regarding the underlying infrastructure. In the Private-Cloud cluster, locality- and scheduling-driven decisions, namely staging data on local disks and pinning services to fixed nodes, consistently reduce both mean runtime and observed variance, in line with expectations. These results are consistent with improved data locality and more stable task placement being associated with lower variability in storage access paths and scheduling behaviour during distributed query execution. In contrast, warm cache and reduced co-tenant load exhibit less intuitive behaviour. Warm cache lowers tail-end variance while increasing typical variability, which may indicate that reducing reliance on storage access lessens the most extreme slowdowns, after which run-to-run differences in task scheduling, network exchange, and in-memory processing become more prominent. Reduced co-tenant load, meanwhile, decreases variance yet increases mean runtime, suggesting that even when competition for shared CPU, network, and storage resources is reduced, end-to-end throughput does not necessarily improve, despite repeated executions becoming more consistent. 

On GCP, introducing similar deployment decisions systematically worsens both runtime and variance. This behaviour suggests that deployment strategies do not transfer directly between environments, likely due to differences in infrastructure abstraction and resource management. For example, cloud platforms may already optimise data locality and resource allocation internally, meaning that manual interventions such as warm caching or alternative storage layers can introduce additional overheads. Arguably, the most interesting finding comes from the external metastore, which is associated with substantial variance despite being provider-managed, suggesting that metadata access itself can contribute to latency variation beyond the query engine and storage layer.

\section{Study~3: Impact on QPP}
\label{Study3}
After establishing that runtime variance is a general challenge for lakehouses, this study examines how the variance regime of runtime labels impacts the accuracy of QPP models.

\begin{objectivebox}[before skip=1.2em, after skip=1.1em]
Analyse how collecting runtime training labels from higher and lower variance lakehouse regimes influences the accuracy of state-of-the-art runtime QPP models.
\end{objectivebox}

\subsection{Methodology}
\looseness=-1We investigate the impact of runtime variance on QPP models by comparing their accuracy under two contrasting variance regimes. For a fixed query workload and model architecture, we consider two variance regimes: a (Higher-Variance) baseline and a Lower-Variance setting, identified as a lakehouse deployment with a lower Avg. Med. and P99 CV. For each deployment, we train a separate QPP model on runtimes from a single pass over the training queries. We then validate each model on the same held-out set of query IDs, with runtime labels taken from the corresponding lakehouse deployment. 

We evaluate two QPP models: the zero-shot cost \enquote{GNN} proposed by Hilprecht and Binnig~\cite{Hilprecht2022a} and a Random Forest \enquote{RF} regressor over SQL embeddings generated by XiYanSQL-QwenCoder-32B~\cite{liu2025xiyan}. We consider three database schemas at SF~1000: TPC-DS, SSB, and JOB. For each workload, we generate 5{,}000 queries using the query generator and tooling released by~\cite{Hilprecht2022a}. Using the Private-Cloud cluster, we execute these queries on the Baseline and \textit{Local disk storage} lakehouse deployments from Study~2. We partition each workload into 80/20 train/test splits, train models on log-transformed runtimes, and select hyperparameters based on validation performance. To understand how relative error scales with variance, given the observed runtime $y_q$ and predicted runtime $\hat{y}_q$ for query $q$, we compute QError as:

\begin{equation}
\mathrm{QError}_q \;=\; 
\max\!\left(\frac{\hat{y}_q}{y_q},\, \frac{y_q}{\hat{y}_q}\right).
\end{equation}

We report mean absolute error (MAE), along with the median (Med.) and 99th percentile (P99) QError.

\begin{figure*}[t]
    \centering
    \begin{subfigure}{\linewidth}
  \centering
  \includegraphics[width=0.95\linewidth]{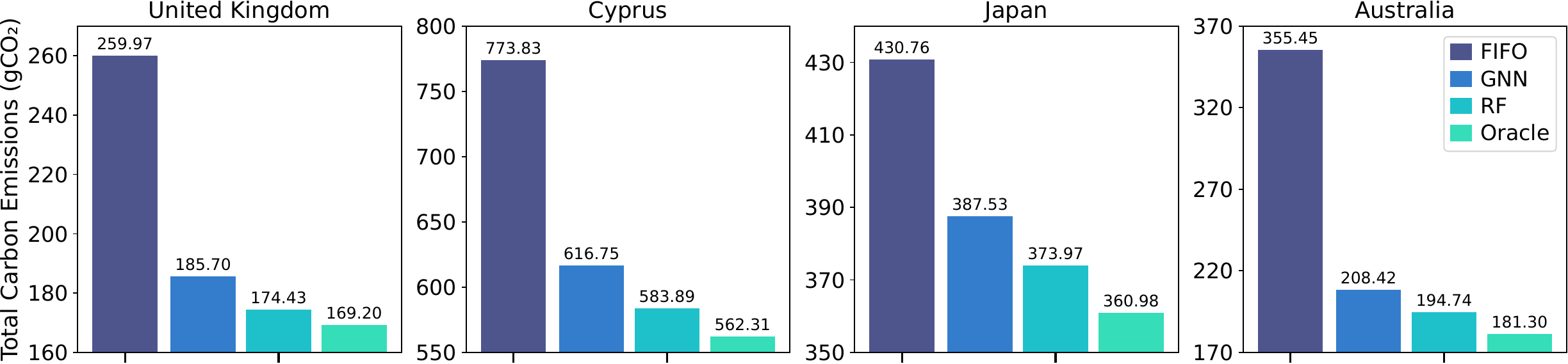}
  \caption{Higher-Variance lakehouse deployments.}\vspace{0.2em}
  \label{fig:total-emissions-baseline}
    \end{subfigure}
    \begin{subfigure}{\linewidth}
  \centering
  \includegraphics[width=0.95\linewidth]{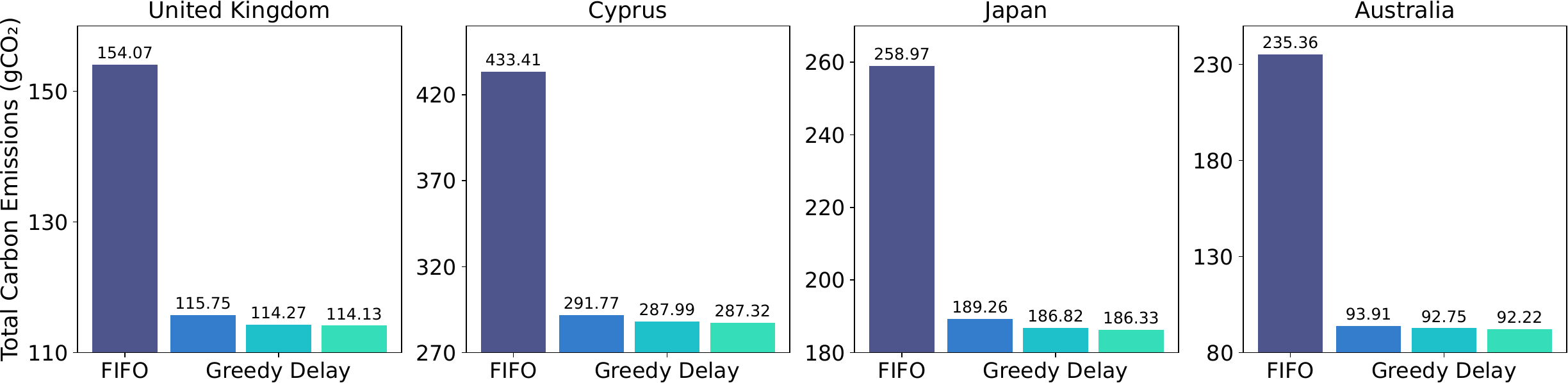}
  \caption{Lower-Variance lakehouse deployments.}
  \label{fig:total-emissions-reduced}
    \end{subfigure}
    \setlength{\abovecaptionskip}{-1em}
    \caption{Total carbon emissions across lakehouses, schedulers, and CI signals.}\vspace{-0.1em}
    \label{fig:total carbon}
\end{figure*}

\subsection{Results}\label{sec:resultsC}

\looseness=-1Table~\ref{tab:qpp-variance} reports the prediction errors for the GNN and RF models trained under the Higher- and Lower-Variance regimes on the Private-Cloud cluster. Across both models, TPC-DS and SSB exhibit larger absolute errors and tails than JOB. Under Higher-Variance lakehouses, both models show large tail errors on TPC-DS and SSB, whereas JOB remains comparatively stable. Moving to the Lower-Variance regime consistently reduces the magnitude of errors for TPC-DS and SSB in terms of MAE and P99 QError for both models, indicating improved tail prediction accuracy. Median QError follows the same direction for the GNN, while for the RF it is less consistent. 

\begin{table}[h]
\centering
\caption{Relative Change in QPP Error}\vspace{-0.5em}
\renewcommand{\arraystretch}{1}
\label{tab:qpp-variance-rel}
\begin{tabular}{cllll
}
\toprule
Dataset & Model & \makecell{$\Delta$MAE\\(\%)} & \makecell{$\Delta$Med QErr\\(\%)} & \makecell{$\Delta$P99 QErr\\(\%)} \\
\midrule
TPC-DS & GNN & $-~49.45$ & $-~32.89$ & $-~57.11$ \\
TPC-DS & RF  & $-~37.27$ & $+~51.41$ & $-~21.19$ \\
SSB    & GNN & \boldmath$-~79.51$\unboldmath & \boldmath$-~47.64$\unboldmath & \boldmath$-~88.24$\unboldmath \\
SSB    & RF  & $-~74.62$ & $-~15.54$ & $-~73.12$ \\
JOB    & GNN & $-~6.13$  & $-~3.91$  & $-~15.52$ \\
JOB    & RF  & $-~2.79$  & $+~2.46$  & $+~0.55$ \\
\bottomrule
\end{tabular}
\end{table}

Table~\ref{tab:qpp-variance-rel} summarises the relative change in error when training on Lower-Variance runtimes. For the GNN, Lower-Variance training yields large improvements on TPC-DS and SSB, reducing MAE by 49.45\% and 79.51\% and reducing P99 QError by 57.11\% and 88.24\%, respectively. For JOB, changes are modest, with MAE and median QError shifting by only a few percent and P99 QError improving by 15.52\%. The RF model shows a similar pattern, with improvements on SSB and moderate gains on TPC-DS in MAE and P99 QError. However, the RF median QError on TPC-DS increases by 51.41\% under Lower-Variance training, and JOB shows negligible changes with slight increases in median and P99 QError.

\begin{takeawaybox}[before skip=1.2em, after skip=1.1em]
Training QPP models on lower-variance lakehouse executions substantially reduces prediction error, lowering MAE by up to 79\% and P99 QError by up to 88\%.
\end{takeawaybox}
Our results show that variance appears as label noise in runtime QPP training data, thereby weakening the learned relationship between query characteristics and observed runtimes. Training models on runtimes from the Lower-Variance Private-Cloud deployment leads to a significant decrease in prediction error for the decision-support workloads TPC-DS and SSB, particularly in reducing tail-end runtime prediction errors. This behaviour may reflect differences in workload characteristics. TPC-DS and SSB contain complex analytical queries dominated by scans, aggregations, and filtering, which may make their runtimes more sensitive to execution variability. By contrast, for JOB, we believe prediction difficulty is strongly tied to the join structures, limiting the potential improvement from reducing runtime variance. The largest improvements are observed in the error tail, suggesting that lower-variance training predominantly suppresses large prediction errors rather than uniformly improving all aspects of a given model's accuracy. The less consistent median-QError behaviour of the RF model further indicates that reducing variance in the training targets does not remove all sources of prediction error. For JOB, baseline models already achieve low absolute error, limiting the potential for further improvement through Lower-Variance training. These findings indicate that variance reduction is most beneficial when label noise is a major contributor to prediction error.

\section{Study~4: Low-Carbon Scheduling}
\label{Study4}
This section presents low-carbon scheduling as a case study to examine how variance in lakehouse performance can impact prediction-driven resource management decisions.

\begin{objectivebox}[before skip=1.3em, after skip=1.3em]
Quantify the impact of runtime variance on low-carbon scheduling by comparing carbon emissions produced when schedules are computed using QPP-predicted runtimes under different variance regimes and CI signals.
\end{objectivebox}

\subsection{Methodology}
\looseness=-1We simulate a scheduler that employs a \enquote{Greedy} low-carbon policy that relies on forecasted runtimes. Using predictions $\hat{y}_q$ for the TPC-DS test workload from the GNN and RF QPP models of Study~3, we compare the carbon emissions produced under the Lower- and Higher-Variance regimes. For our baseline, we consider emissions when predicted runtimes are replaced with observed \enquote{Oracle} runtimes $y_q$. To validate the effectiveness of our Greedy low-carbon scheduling heuristic, we also evaluate the emissions resulting from a rudimentary first-in-first-out \enquote{FIFO} scheduler.

To implement our Greedy scheduler, we consider a single-slot, non-preemptive cluster that schedules a batch workload of $N$ analytical queries released at time $t_0$. For all unscheduled queries in the batch, we identify the time $s_q$ to execute the next query $q$ so that the predicted execution window $[s_q,\, s_q + \hat{y}_q]$ yields the least carbon emissions within a 12-hour decision window. Given the resulting schedule, we calculate the execution window of query $q$ as $[s_q, s_q + y_q]$, and estimate carbon emissions by finding the product of the carbon intensity $\mathrm{CI}(t)$ and power consumption $P(t)$ during execution:

\begin{equation}
\!\!\!C_q
 \; = \int \mathrm{CI}(t)\,P(t)\,dt
\;\;\approx \!\!\!\!\!\!
\sum_{k \in [s_q,\,s_q + y_q]} \!\!\!\!\!\!\!\!\mathrm{CI}[k]\,P[k]\,\Delta
\end{equation}
Carbon-intensity traces are taken from Electricity Maps for four locations (The United Kingdom, Cyprus, Japan, Australia) over a 5-day horizon (from 9 to 14 November 2024). As we focus on relative differences between schedulers and variance regimes, we assume each lakehouse draws constant power $P = 0.15~\text{kW}$ while executing queries and ignore idle power consumption. We report the total carbon emissions produced for each scheduled workload and quantify the impact of variance by presenting each scheduler’s relative overhead versus the Oracle under the two lakehouse-variance regimes.

\subsection{Results}\label{sec:resultsD}
\begin{figure}[!t]
\centering
\includegraphics[width=\linewidth]{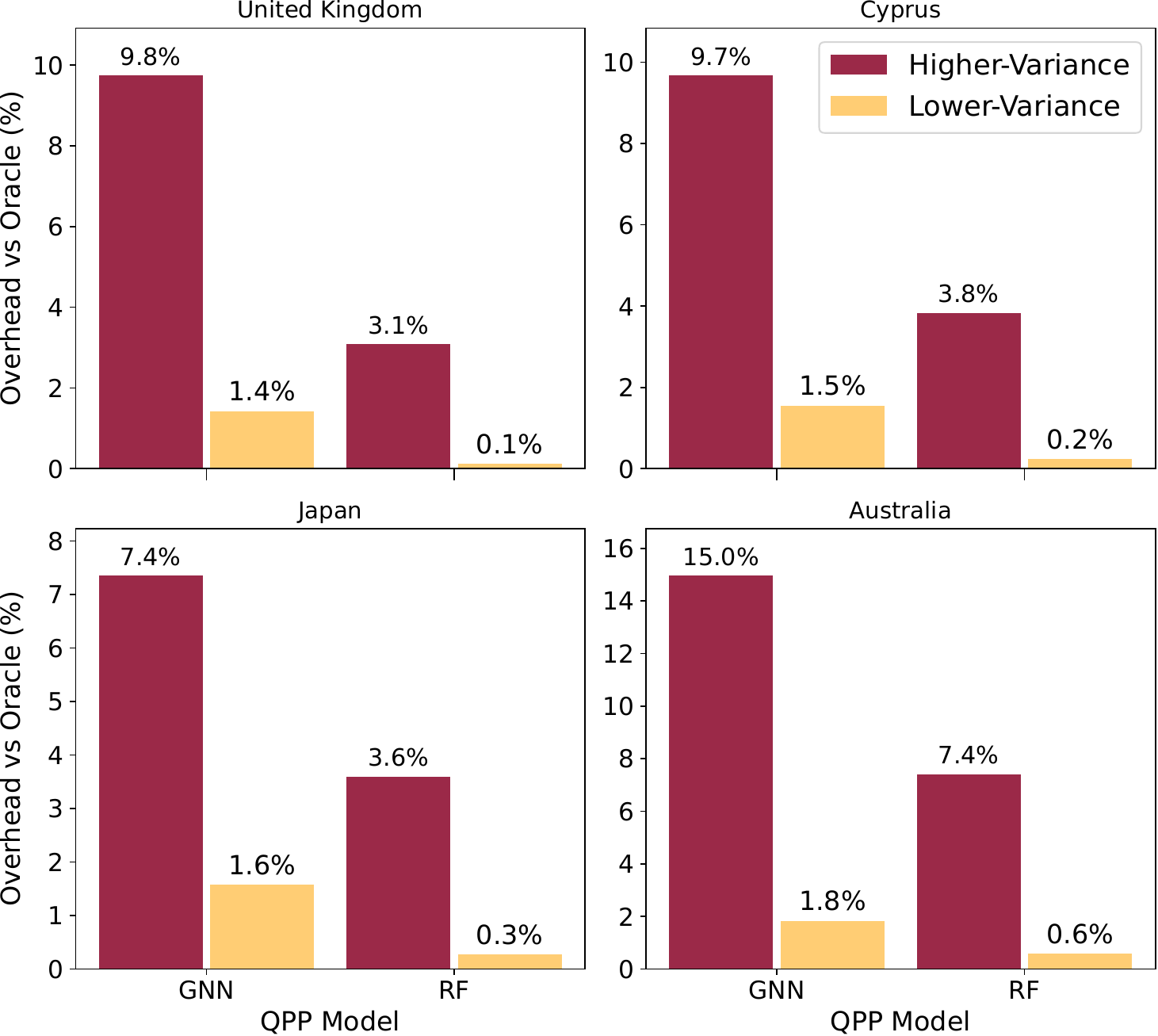}
\caption{The relative difference in carbon emissions compared to the oracle schedule: Higher- and Lower-Variance lakehouses.}\vspace{-1em}
\label{fig:carbon-overhead}
\end{figure}

\looseness=-1Figure~\ref{fig:total carbon} presents total carbon emissions for schedulers across the four CI regions under both Higher- and Lower-Variance regimes. In all regions and regimes, FIFO yields the highest emissions, whereas adopting low-carbon scheduling notably reduces emissions towards those observed for the respective oracle. For instance, when viewing the emissions of the United Kingdom setting under the Higher-Variance setting, emissions decrease from 259.97~gCO\textsubscript{2} with FIFO to roughly 174--186~gCO\textsubscript{2} with Greedy scheduling. Interestingly, we observe that across all locations, schedules derived from the RF QPP model are consistently lower than those from the GNN model. This behaviour is consistent with the lower tail prediction errors observed for the RF model in Study~3, as scheduling decisions are particularly sensitive to large runtime prediction errors that cause queries to overrun into higher carbon-intensity periods. Figure~\ref{fig:carbon-overhead} presents these results as relative overhead compared to the Oracle scheduler. Under Higher-Variance conditions, all Greedy Delay schedulers exhibit overhead, ranging from 15.0\% to 7.4\% for the GNN model and from 7.4\% to 3.1\% for the RF model, with the highest overhead occurring in Australia. When Lower-Variance runtime predictions are used, overheads consistently decrease across all CI zones. For example, in Japan, the GNN overhead declines from 7.4\% to 1.6\%, and the RF overhead from 3.6\% to 0.3\%.

\begin{takeawaybox}[before skip=1em, after skip=1em]
Prediction errors resulting from runtime variance impact low-carbon scheduling: In higher-variance lakehouse regimes, prediction-based schedulers produce greater carbon emissions, whereas in lower-variance regimes, these schedules are closer to oracle emissions.
\end{takeawaybox}

Results from the case study illustrate how runtime variance affects query orchestration decisions and, consequently, the emissions observed under the evaluated CI traces. Across different regions, most low-carbon schedulers outperform FIFO. However, under high-variance conditions, they still experience significant overheads compared to the oracle scheduler, particularly in regions with fluctuating CI signals such as the United Kingdom and Cyprus. In these cases, skewed QPP runtime estimates, resulting from variance in training labels, lead to suboptimal scheduling decisions. When variance is reduced and QPP prediction tails narrow, the overhead relative to the oracle scheduler decreases to only a few percent.

\subsection{Uncertainty-Aware Query Scheduling}

To offer preliminary insight into whether low-carbon query scheduling can benefit from directly accounting for runtime variance, we investigate an uncertainty-aware scheduling heuristic that adopts conservative query placement. We hypothesise that by padding scheduled windows for incoming queries, the likelihood that queries overrun their assigned windows and spill into periods of higher CI may be reduced, thereby yielding lower emissions across the entire schedule. Using the Higher-Variance lakehouse setting, across the four CI regions and for both the GNN and RF models, we increase the scheduled window for each query by increments of 10\% from 0\% to 100\% (guided by the observation that the P99 CV for queries under this setting is 98.65\%) and compare the resulting carbon emissions against their unadjusted schedules.

\begin{figure}[!h]
\centering
\includegraphics[width=\linewidth]{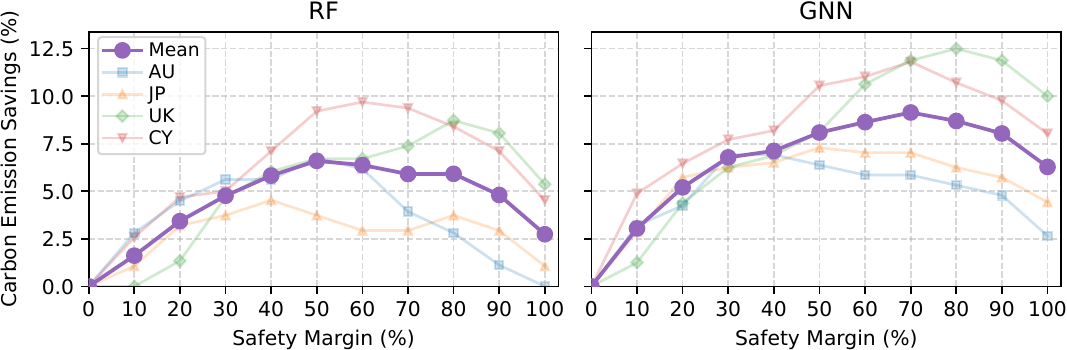}
\caption{Average carbon emission savings across CI regions as scheduled query windows are increased from 0\% to 100\%.}\vspace{-0.6em}
\label{fig:padding-sensitivity}
\end{figure}

\looseness=-1Figure~\ref{fig:padding-sensitivity} shows that introducing safety margins to scheduled query windows can reduce carbon emissions relative to unadjusted schedules across both QPP models and all four CI regions. For the RF model, mean savings increase up to 50\%, after which gains begin to diminish. For the GNN model, the mean trend continues to improve up to 70\% before savings begin to taper. Overall, the results are consistent with our hypothesis that accounting for variance in prediction-driven orchestration tasks is beneficial, although the tapering at higher margins here suggests that overprovisioning can offset potential gains. By examining the region-specific curves, we observe that the margin associated with the largest gains is not uniform across CI regions. This suggests that, rather than identifying a universal policy across regions and prediction models, runtime uncertainty may need to be accounted for at finer granularity, such as across model-region pairs or even individual queries.

\section{Discussion}\label{sec:discussion}
This section discusses the implications of addressing runtime variance and potential threats to validity. 

\subsection{Implications of the Results}
Our investigation into runtime variance in lakehouses demonstrates that QPP and prediction-based workload orchestration are highly dependent on the certainty of query runtimes. To offer guidance, we discuss approaches to address variance:

\textbf{First,} runtime variance can be reduced through considerate lakehouse deployment choices. Our factor analysis shows the clearest reductions occur when execution improves data locality through local copies, uses consistent node placement through pinning, and limits exposure to resource sharing. In practice, however, these conditions require trade-offs that may be feasible in controlled experiments but difficult to sustain in production deployments.

\textbf{Second,} variance may be addressable by explicitly accounting for primary sources within runtime prediction modelling. For instance, QPP models could be augmented with features that proxy these sources such as network performance or co-tenant load. This direction is appealing but is inherently limited by what can be observed and relied upon at prediction time, as dominant sources of variance may be hidden by platform services, such as data locality, or fluctuate over time, due to interference with co-located load.

\looseness-1\textbf{Third,} variance could be treated as a signal by propagating runtime uncertainty into downstream decision-making, rather than relying solely on point estimates. Our uncertainty-aware scheduling study provides an initial demonstration of how this idea can be applied in practice. More broadly, our results suggest that prediction-driven workload orchestration could benefit from explicitly reasoning about the uncertainty and potential execution risk associated with individual queries. Overall, we view this as the most promising direction, offering benefits even when variance cannot be fully eliminated or sources of variance cannot reliably be factored in at prediction time.

\subsection{Threats to Validity}
\textbf{Lakehouse design:}
We fix the query engine and table format to isolate runtime variance from differences in planning, execution, and table semantics. Although absolute performance may vary across lakehouse deployments, the mechanisms we study arise from running lakehouse queries on cloud infrastructure and in a distributed manner. Our results should therefore be read as evidence for a representative open-source lakehouse stack, while the magnitude of variance may differ across engines, table formats, and storage systems.

\textbf{Factor coverage:}
Study~2 targets a selected set of factors that commonly affect performance predictability in distributed settings, including data locality, system composition, and co-located load. We do not attempt to exhaustively identify every possible variance driver, nor do we test possible combinations of factors. For example, we do not explore layout and format choices (e.g., file sizing, compression, indexing), which may also influence variance. We omit these because a comprehensive exploration would substantially expand the search space.

\textbf{QPP models:}
\looseness=-1We evaluate two representative QPP approaches, a GNN-based estimator and a Random Forest regressor over learned embeddings, rather than examine an exhaustive set of model families or feature encodings. While prediction errors may differ across other models, we observe consistent results across workloads: Lower-Variance lakehouses yield lower MAE and smaller QError tails for QPP models. We do not repeat full end-to-end QPP training and evaluation across multiple independent runs due to the high collection cost, and instead study variance through controlled deployment settings shown in Study~2 to induce systematic changes.

\textbf{Scheduler design:}
\looseness=-1We acknowledge that our scheduling case study abstracts away many production constraints (e.g., cost constraints, service level targets, power consumption), which may limit practical insight. However, this decision was made to isolate the interactions between runtime variance, prediction error, and CI signals. 

\section{Conclusion}\label{sec:conclusion}
\looseness=-2In this paper, we investigated how runtime variance manifests in cloud deployments of distributed data lakehouses and how it propagates into query performance prediction and low-carbon scheduling. Across four public and private cloud setups, we showed that query runtimes exhibit persistent variance, irrespective of lakehouse scale. A controlled factor analysis revealed that deployment choices such as local storage and pinning nodes can substantially reduce both mean runtimes and variance, whereas external services and co-tenant load amplify uncertainty. Building on this, we demonstrated that training QPP models on reduced-variance labels yields smaller prediction errors, and that these improvements translate into improved prediction-based low-carbon scheduling with reduced emissions relative to an oracle. Finally, we provided preliminary evidence that explicitly accounting for runtime uncertainty during scheduling can improve prediction-driven scheduling decisions and yield further reductions in emissions.

Overall, our results show that runtime variance is a highly relevant concern for cloud-based lakehouse deployments and directly impairs the performance of QPP and downstream query scheduling. Looking ahead, we plan to extend this analysis to a broader range of engines, table formats, and workload types, and to explore uncertainty-aware scheduling policies that explicitly account for performance variance rather than treating it as negligible noise.

\section*{Acknowledgment}
\noindent We thank Google Cloud Platform for providing research credits to support this work. We also gratefully acknowledge Electricity Maps as a source of historical electricity grid data.

\bibliographystyle{IEEEtran}
\bibliography{refs}

\clearpage

\end{document}